\documentclass[runningheads]{llncs}
\usepackage{graphicx}
\usepackage{hyperref}
\usepackage{comment}

\begin{document}
\title{A trustable and interoperable decentralized solution for citizen-centric and cross-border eGovernance: A conceptual approach}
\titlerunning{A decentralized solution for citizen-centric eGovernance}
%
\author{George Domalis\inst{1,3} \and 
Nikos Karacapilidis\inst{2} \and
Dimitris Tsakalidis\inst{1} \and
Anastasios Giannaros\inst{3}}
\authorrunning{Domalis et al.}
%
\institute{Novelcore, Greece \and
IMIS Lab, MEAD, University of Patras, Greece \and
Computer Engineering and Informatics Department, University of Patras, Greece
\\
\email{\{domalis,tsakalidis\}@novelcore.eu, \{karacap,up1070374\}@upatras.gr}}
\maketitle              
\begin{abstract}
Aiming to support a cross-sector and cross-border eGovernance paradigm for sharing common public services, this paper introduces an AI-enhanced solution that enables beneficiaries to participate in a decentralized network for effective big data exchange and service delivery that promotes the once-only priority and is by design digital, efficient, cost-effective, interoperable and secure. The solution comprises (i) a reliable and efficient decentralized mechanism for data sharing, capable of addressing the complexity of the processes and their high demand of resources; (ii) an ecosystem for delivering mobile services tailored to the needs of stakeholders; (iii) a single sign-on Wallet mechanism to manage transactions with multiple services; and (iv) an intercommunication layer, responsible for the secure exchange of information among existing eGovernment systems with newly developed ones. An indicative application scenario showcases the potential of our approach.

\end{abstract}
%
%
\section{Introduction}\label{introduction}
The rapid growth of Internet technologies, mobile communications, cloud infrastructures and distributed applications have brought an unprecedented impact to all spheres of the society and a great potential towards the establishment of novel eGovernance models \cite{WimmerVialePereiraRonzhynSpitzer2020}. These models should deploy core values such as improved public services and administrative efficiency, open government capabilities, improved ethical behavior and professionalism, improved trust and confidence in governmental transactions \cite{TWIZEYIMANA2019167}. Towards the modernization of their services and the reduction of the associated bureaucracy, public administrations need to transform their back-offices, upgrade their existing internal processes and services, and provide privacy-preserving and secure solutions. It is necessary to leverage key digital enablers, such as open services and technical building blocks (eID, eSignature, eProcurement, eDelivery and eInvoice), shared and reusable solutions based on agreed standards and specifications (Single Digital Gateway), as well as common interoperability practices (e.g. European Interoperability Framework). This leads to the upgrade of services that enable cross-border data sharing among public administrations, businesses and citizens. 

Governance models, in general, do not adopt a citizen-centric paradigm \cite{Ghareeb2019}; they do not take into account citizens’ needs and expectations of new services, often excluding them from operational and decision making processes. Moreover, documentation exchanges, processes and contact points do not function as a whole; they are rather dispersed and not sufficiently inter-connected among countries and organizations. Transparency and accountability are additional aspects of major importance towards building a good and fair governance model \cite{Bertot2010}. Admittedly, governments that employ models to make information sharing and decision-making processes transparent improve the principle of accountability and augment the participation of citizens and other stakeholders in related actions \cite{HARRISON2014513}. The corresponding digital transformation of public services can reduce administrative burdens, enhance productivity of governments, minimizing at the same time all the extra cost of traditional means to increase capacity, and ultimately improve the overall quality of interactions with and within public administrations \cite{Androutsopoulou2019}. 

Taking into account the above issues, this paper introduces a transparent, cross-border and citizen-centric eGovernance model for public administration services, which automates the processes and safeguards the integrity of interactions among citizens, businesses and public authorities. By taking advantage of emerging ICT technologies, such as Peer-to-Peer (P2P) networks, Distributed Ledger Technologies (DLTs) and smart data structures, we deploy a public distributed infrastructure, based on the InterPlanetary File System (IPFS) and a distributed ledger. This solution is based on a single sign-on Wallet mechanism that interconnects distinct decentralized applications (dApps) responsible for ID authentication, document sharing, information exchange and transactions validation, enabling a single point of access to information. The proposed solution is fully in line with the Government 3.0 paradigm, in that it meaningfully integrates a diverse set of disruptive and established ICTs \cite{unknownTerzi}.

The remainder of this paper is organized as follows: \hyperref[section2]{Section 2} is devoted to the presentation of the underlying technologies employed in our approach. The proposed digital transformation model, enabled through an architecture incorporating a series of prominent technologies, is presented in \hyperref[section3]{Section 3}. Particular emphasis is given to the inclusion of data governance and knowledge management services to best facilitate and eventually reduce lengthy, cumbersome and repetitive bureaucratic eGovernment transactions. \hyperref[section4]{Section 4} validates the potential of our approach through a representative application scenario. Finally, \hyperref[section5]{Section 5} outlines concluding remarks and future research directions.

\section{Underlying Technologies}\label{section2}
\subsubsection{Interplanetary File System.}\label{subsection2.1}
It is a P2P distributed file sharing system that seeks to connect all computing devices with the same file system by providing a high throughput content-addressing block storage model. IPFS distributes files across the network. Each file is addressed by its cryptographic hash based on its content, rather than its location as in traditional centralized systems where a single server hosts many files and information has to be fetched by accessing this server. This characteristic renders IPFS an ideal data storage solution for eGovernment services, where security and transparency are of utmost importance, since there is no single point of failure. One of the main content routing systems of the IPFS architecture is the Distributed Hash Table (DHT), which allows key-based lookup in a fully decentralized manner. IPFS leverages a DHT system, in that all IPFS nodes "advertise" content items stored in the DHT and this results in a distributed dictionary used for looking up content. Various applications integrating IPFS with ledger and blockchain technologies have already been reported in the literature for transactions recording \cite{10.1145/3409934.3409948} and secure file sharing with decentralized user authentication, access control and group key management mechanisms \cite{8540048,8835937}. In the approach described in the next section, we leverage IPFS to provide increased capacity for managing large datasets in a decentralized manner and complement the throughput limitations of distributed ledger technologies.
\vspace{-3mm}
\subsubsection{Distributed Ledger.} \label{subsection2.3}
It is a distributed database architecture that records transactions on a P2P network and enables multiple members to maintain their own identical copy of a shared ledger without the need for validation from a central entity. Transaction data are scattered among multiple nodes using the P2P protocol principles, and are synchronized at the same time in all nodes. A public distributed ledger is characterized by an open unprotected environment with millions of participants, most of which have limited computational power and bandwidth, while most of the power is in the hands of a small fraction of the participating nodes. Thus, there is a significant risk of a "majority attack", in which a few nodes can dictate the choice of transactions. For eGovernance purposes, the design of a distributed ledger requires a comprehensive approach taking into account diverse aspects such as intermediate scale, high processing rate and low completion time with moderate energy consumption, unique attack model, and utilization of the underlying data structures. 
 
\vspace{-3mm}
\subsubsection{Smart contracts.}\label{subsection2.5}
They are decentralized, trusted computer programs stored on a blockchain that are automatically executed when predetermined terms and conditions are met. They facilitate, verify, or enforce documents and actions according to the terms of a contract or an agreement between two parties (i.e. agreements between eGovernance operators of two countries and end-users) that consist of a set of rules dictating a reaction when specific actions occur \cite{8500488}. This set of rules is deployed on blockchain to ensure decentralized, transparent and secure characteristics. Upon meeting predefined conditions, a smart contract is executed automatically, making it independent of any central entity. Shields et al. \cite{OShields2017SmartCL} discuss the use of smart contracts for legal agreements and conclude that smart contracts will benefit from the legal precedent established in the electronic marketplace. In our approach, smart contracts are employed to manage the automatic execution of policies for the proposed services.
\vspace{-3mm}
\subsubsection{Decentralized Applications.}\label{subsection2.4}
They are composed of distributed entities that directly interact with each other and make local autonomous decisions in the absence of a centralized coordinating authority. According to Raval \cite{10.5555/3074145}, a dApp is characterized by four features: (i) open source, (ii) internal currency, (iii) decentralized consensus, and (iv) no central point of failure. Bittorrent \cite{cohen2003} was the initial dApp, enabling users to connect and exchange files. Soon afterwards, blockchain technology was introduced to manage the decentralization and enable the immutability of data. In the context of eGovernment 3.0, dApps can be leveraged to deliver diverse services such as ID authentication, document sharing, information exchange and transactions validation.
\vspace{-3mm}
\subsubsection{Smart data structures.}\label{subsection2.6}
The incorporation of Machine Learning techniques to explore causal relations among Big Data \cite{somani2017} in eGovernment systems is often deterred by interoperability inefficiencies \cite{10.1145/3428502.3428536}. \textit{Smart data structures} \cite{Eastep2011} are a new class of parallel data structures that leverage online Machine Learning and self-aware computing principles to tune themselves automatically. They can replace existing index structures with other types of models, including deep learning models, referred to as learned indexes \cite{10.1145/3318464.3389711,10.1145/3183713.3196909,10.1145/3332466.3374547}. Recent works present preliminary outcomes of the conceptual and methodological aspects of semantic annotation of data and models, which enable a high standard of interoperability of information \cite{Villa2017} and showcase how multi-input deep neural networks can detect semantic types \cite{10.1145/3292500.3330993}. We utilize smart data structures to identify and effectively transform data schemas and interconnect existing centralized systems with our decentralized solution.
\vspace{-3mm}
\subsubsection{Single sign-on.}\label{subsection2.7}
It is an authentication mechanism that enables the use of a unitary security credential to access related, but independent, software systems or applications \cite{koundinya2020review}. It enables simple username and password management, improved identity protection, increased speed and reduction of security risks. It also includes functionalities such as password grant (sign-in directly on the web), authorization code grant (user authorizes third-party), implicit grant (third-party web app sign-in), web services API that can effectively authenticate requests, and seamless user authorization experience on client-side technology. Various types of schemas exist based on (i) the type of infrastructure, (ii) the system architecture, (iii) the credential forms (token, certificate), and (iv) the protocols used. It is a critical part of complex environments, where multiple services from various providers are hosted. While single sign-on has been mainly used in mobile and Internet of Things applications, its integration with distributed file systems still remains a challenge.

\section{The proposed solution}\label{section3}
\subsection{Research methodology}\label{subsection3.1}
For the development of the proposed open and cross-border eGovernance model, we have adopted the \textit{design science paradigm} \cite{10.5555/1859261}, which aims to extend the boundaries of human and organizational capabilities by creating new and innovative artifacts, especially for the information technologies domain. For our purposes, we have used the specific design science research methodology proposed by Peffers et al. \cite{peffers2008design} for the domain of Information Systems (IS) research, which includes the following stages: identify problem and motivation, define objectives of a solution, design and development, demonstration, evaluation and communication.

Furthermore, we have combined the above paradigm with that of \textit{action research}, which aims to contribute both to the practical concerns of people in an immediate problematic situation and to the goals of social science by joint collaboration within a mutually acceptable ethical framework. It has been recognized that action research can be quite important in the IS domain, as it can contribute to improving its practical relevance \cite{Iivari2009ActionRA}. In particular, it enables the design, implementation and evaluation of ICT-based actions/changes in organizations, which address specific problems and needs that are of high interest for practitioners, and at the same time create scientific knowledge that is of high interest for the researchers. The complementarity between these two research paradigms, as well as the great potential of integrating them, have been extensively discussed in the literature; both paradigms aim to directly intervene in real-world domains and introduce meaningful changes in them.

In particular, to address the issues elaborated in this work, we cooperated with two Greek government agencies (the Ministry of Digital Governance and a big local municipality) and two organisations with long experience in the development of novel software solutions for eGovernment. Involving their most experienced staff, we organized three workshops of 2 hours duration each. In these workshops, we followed a qualitative approach (in-depth discussions) to collect relevant information and accordingly shape the foreseen services. Based on the information collected, we designed the solution presented below. 

\subsection{Our approach}\label{subsection3.2}

\begin{figure}
\centering
\includegraphics[width=1\textwidth]{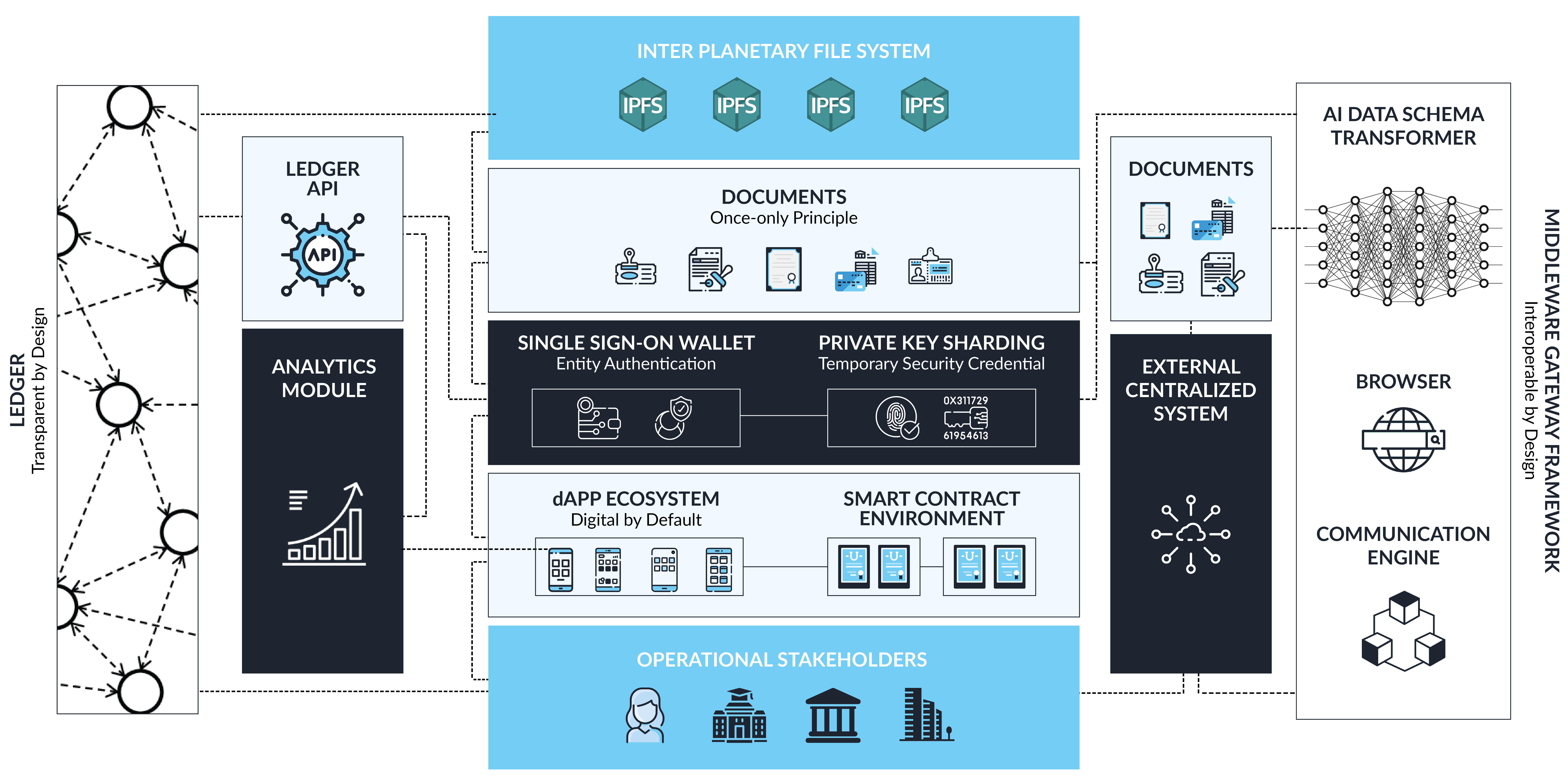}
\caption{The architecture of the proposed solution.} \label{fig1}
\end{figure}

We propose a novel eGovernance model that creates new digital governance pathways through the integration of emerging technologies and breakthrough cross-sector services. We deploy decentralized applications (dApps) to deliver efficient, reliable and secure data sharing, auditing mechanisms and communication channels for the eGovernment sector. Our overall approach is digital by default, transparent and interoperable by design, and fully adheres to the once only priority. All the individual modules are designed to be built on top of the IPFS and a distributed ledger, as illustrated in \hyperref[fig1]{Fig. 1}. This extended combination of distributed technologies and infrastructures constitutes the backbone of our solution, which is capable of efficiently addressing the complexity of the processes, providing at the same time the security, trustworthiness, immutability and auditability required by contemporary public services.

By combining functionalities enabled by \textit{DLTs} and \textit{IPFS}, our approach allows users to control their data without compromising security or limiting third-parties to provide personalized services. IPFS has specific features that remedy the performance issues of dApps, improving their performance through an ad-hoc engagement of existing computational and storage resources. These features include: (i) content indexing, (ii) hash lookup, (iii) distributed naming system (IPNS, similar to DNS), (iv) persistence and clustering of data that reduce latency, (v) decentralized archiving, and (vi) compliance with privacy regulations.

The public ledger stores the users' digital identities, access consent logs, and selected authentication transactions. It co-supports the public sharing resource infrastructure, providing the IPFS with advanced operational and technological capabilities, starting from security and privacy, up to immutability of transactions. It is located at the heart of the network to monitor and continuously record every approved interaction among the users' nodes. One of the major challenges is how to overcome the public nature of the ledger to ensure the security, privacy and anonymity of information. Towards this direction, we propose a combination of private and public keys to exchange information among end-users and services. In this way, a service does not observe raw data, but instead it runs its computations directly on the network and obtains the final results only after the consent of the user. We use DLTs since, from the data safety, authenticity and non-repudiation point of view, they provide an easily accessible and immutable history of all contract-related data, adequate for building applications with trust, accountability, and transparency. 

The \textit{dApps} are used to deliver dedicated services to users. They comprise smart contracts and are hosted in the nodes of the distributed network. The overall performance and storage capacity are increased through the participation of new nodes joining the network. Our approach renders legal and regulatory decisions simple, since law and regulations are programmed by smart contracts in the network and are enforced automatically, while the ledger acts as a legal evidence for storing such data.

A \textit{single sign-on Wallet} is responsible for managing - with a common person registry - multiple services provided by various dApps, such as document sharing and information exchange, enabling a single sign-on user-centric document repository. This tool allows stakeholders to manage and share their personal and sensitive documents among different application environments in a single management kit, which can function in a fully distributed way, without a single point of failure. Hence, providing resilience and a continuum of service. 

In addition, a middleware layer includes the following modules: (i) a \textit{Secure Gateway Channel on-the-fly} to assure a secure intercommunication among systems, databases, apps, etc; (ii) an \textit{AI Data Schema Transformer}, which is based on well-defined libraries and AI models (i.e. GPT2 \cite{Radford2019}) and encompasses the EU interoperability standards to effectively identify and transform data schemas, models and structures, and enable machine-to-machine communication among different types of systems across the EU. Synthetic data collections that map and simulate real data types (i.e. citizens ID, passport, birth certificate, criminal record, etc.) for different EU countries provide the backbone of this module's perception; (iii) a \textit{Transactions-based Analytics Module} that runs as a back-end service on the network, gathering transaction histories and provide insights to users through a user interface developed and delivered as a dApp; and (iv) a \textit{Browser Service Module} that acts as a search engine and facilitator of the network and other modules (i.e. the dApp ecosystem), providing users with diverse functionalities (locate files, documentation, services, entities, etc.) through user-friendly interfaces. This extended digital service availability enables any physical and/or legal entity (such as public administration, business and citizens) to integrate their own external centralized system in the network, enabling interoperability between users, cross-border and cross-sector organizations.

To effectively satisfy the desired interoperability by design principle, we deploy a machine learning based environment that automatically recognizes data structures in existing centralized systems. Specifically, we apply the notion of smart data structures by employing deep learning techniques to recognize and transform data schemas, data structures and data types. We apply data fusion techniques to meaningfully integrate heterogeneous information from multiple data sources that would otherwise remain uncorrelated and unexploited. To build a highly tuned system tailored to the specific needs of eGovernment services, we identify the data distributions and examine possible optimizations in the index structures to identify data patterns. The key idea is that a model can learn the sort order or structure of lookup keys and use this signal to effectively predict the position or existence of the associated records. 

This holistic framework creates a novel eGovernance mechanism for communication, data sharing and information retrieval among centralized systems and decentralized/distributed applications. Stakeholders are able to use the single sign-on Wallet and the individual modules running on the network, while also having unrestricted access to the dApp ecosystem. This ecosystem hosts different dApps for each distinct service supported by the network, which are custom designed and deployed to meet the needs of the end users. The functionalities and the interdependencies of the dApps are formulated and defined with the use of dedicated smart contracts and user interfaces, hosted under the ecosystem. The dApp ecosystem includes libraries of common deployed smart contracts, enabling their reuse and activating users/developers to build and deploy their own applications. Finally, the distributed file storage architecture gives the opportunity to stakeholders to host their own custom applications in the network, thus contributing to its expansion and scalability.

\section{An application scenario}\label{section4}
According to the Treaty on European Union and Community law, EU citizens have the right to work and live in any EU Member State. However, the process of moving abroad to a foreign country is quite complex in terms of bureaucratic processes. For instance, a Social Security Number (SSN) is required in many EU countries before signing a rental contract. In most cases, the burden of such actions rests solely with the citizens, not in terms of legislation but in terms of complexity of the processes that need to be carried out, let alone the multiple visits people have to make to the relevant public authorities. The following scenario showcases the application of our user-centric solution, which simplifies the bureaucratic processes for citizens, businesses and public administrations.

\subsubsection{Overview.}
Alice, a Greek citizen, finds a vacant job position in a private company in Portugal. She applies for the job and thankfully gets hired. In Portugal, she has to deal with a series of bureaucratic processes (issue an ID card and a SSN, open a bank account, provide evidence of her educational certificates etc.). To obtain a residence title, rent an apartment and open a bank account, she needs to present at least a validated ID documentation, a birth certificate, a nationality certification validated by a Greek Authority and a proof that she works in Portugal, along with the additional information that may be required by the employer. Adopting our solution, Alice is able to request from the Greek Authorities (Ministry of Digital Governance - MoDG) the proof of ID and the required data, validated. At the same time, Alice can remotely request from her formal educational institution (University of Patras - UoP) all the required certificates (diploma etc.).

In turn, MoDG issues the document and Alice gives permission (using a distributed application) to forward it to the respective public authorities in Portugal (Ministry of Justice - MoJ). As soon as this transaction is completed, Alice obtains, has access and is able to securely share her Portuguese SSN through her Wallet. Now, her employer in Portugal can directly get the validated SSN from the Portuguese MoJ, after her approval to register her credentials to their internal payroll system. Furthermore, the HR representative of the company needs a series of legal documents to proceed with the hiring process, including the permanent visa permit. Through her Wallet, Alice is able to provide and/or revise all the required personal documents.

\subsubsection{Added value.}\label{subsection4.3.2}
The contribution of this approach stands on the simplification of the processes and the significant reduction of bureaucracy. According to the current legislation and official procedures, the process to move to a foreign country for a new job is complex. Opening a bank account, renting an apartment or proving educational certification and achievements can be highly demanding in terms of paperwork, since in many cases there are requests that necessitate cross-border exchange of information and associated documents. Through the proposed eGovernance solution, inconsistencies in bureaucratic procedures will be avoided, such as requiring a local bank account before being able to rent an apartment, while at the same time requiring a local address before being able to open a bank account. 

\begin{figure}
\includegraphics[width=\textwidth]{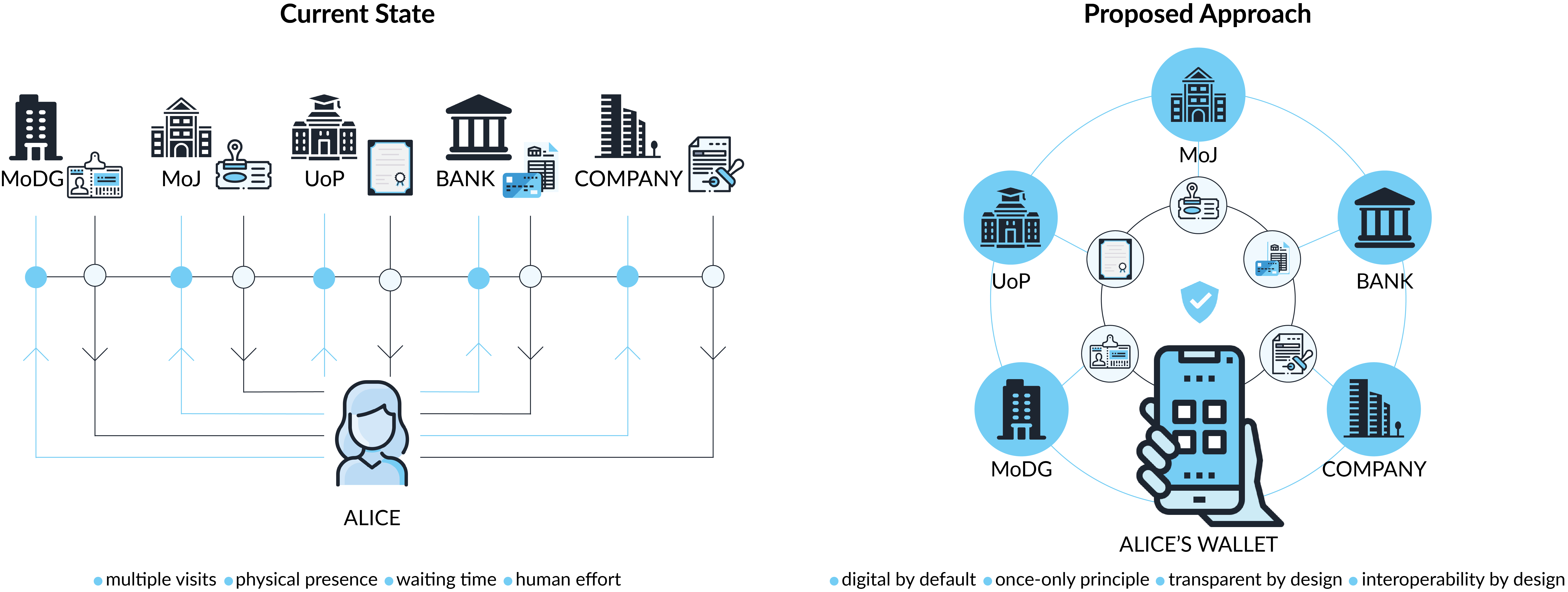}
\caption{Current practice compared to the proposed solution.} \label{fig2}
\end{figure}

Through a set of dApps from the distributed application ecosystem, which adopt a single sign-on Wallet approach and exploit the IPFS and distributed ledger infrastructure, Alice can use her validated digital identity to remotely request, obtain and share the required legal documents and certifications. Upon her permission, all these can be moved directly from the issuing authorities (MoDG, MoJ and UoP) to her new employer and any other potential entity (bank, utilities, etc.). These authorities digitally issue and validate the documentation and instantly push encrypted data into the distributed network, while the transactions among the users are being recorded. Any type of transactions, including requests, notifications and permissions, are monitored and safely stored to protect Alice’s privacy. All transactions are stored in a public ledger to enhance the security of information and eliminate any forgery attempts. Hence, enabling a secure-by-design eGovernment solution. This approach brings forward multiple advantages such as minimizing the chance of forged documentation, enhancing the transparency and security of information exchange in adherence with the relevant legislation, decreasing the overall effort of citizens and reducing significantly the waiting time to carry out the necessary transactions and issue the corresponding documents. 

\section{Conclusions}\label{section5}
This paper has described a novel eGovernance model that aims to make public administrations and public institutions open, efficient and inclusive, providing border-less, digital, personalised and citizen-driven public services. This solution offers citizens and businesses efficient and secure mobile public services and co-creation mechanisms, enabling governments to be extroverted and to preserve trust among public and private entities. The services delivered to citizens contribute to the digital by default principle for government and local authorities. In addition, our approach enables beneficiaries to participate and operate in a by design efficient, cost-effective, secure and cross-border distributed network for data exchange and service delivery. 

The main contribution of this work is the meaningful integration and orchestration of a set of prominent tools and services, which build on state-of-the-art technologies from the areas of distributed computing and artificial intelligence, to address requirements concerning simplification of processes and reduction of bureaucracy in diverse eGovernment transactions. Our approach has been validated and elaborated in close co-operation with three government agencies (Portuguese Ministry of Justice, Greek Ministry of Digital Governance, and Istanbul Metropolitan Municipality), through which a series of rich application scenarios have been sketched and analyzed. While the feedback received from such a first-level validation was positive, and the proposed solution is open and inclusive by design, its application has to carefully consider the information capacity and available resources of each public sector organisation. Moreover, it has to be evaluated through diverse 
usefulness and ease-of-use indicators.

The proposed approach has interesting research and practical implications. With respect to research, it leverages the existing knowledge in the utilization of distributed computing and AI technologies in the public sector, and advances the digital transformation of eGovernment transactions. With respect to practice, our solution deploys a novel digital channel of communication and collaboration between citizens, businesses and governments. It addresses fundamental weaknesses of the existing eGovernment transactions in terms of bureaucracy, complexity, and unnecessary data entry, while also leveraging existing resources and infrastructures. As a last note, we mention that our approach can be applied in several real-life scenarios, such as the identification control in airports, where passengers need to be checked before departing and after reaching their destination. This application can also incorporate the management of the currently elaborated COVID-19 vaccination certificates \cite{KoflerBaylis2020}; the proposed decentralized blockchain ledger enable an immutable and transparent solution, according to which entries can be publicly audited and anonymity is protected. 

\subsubsection{Acknowledgements:} This publication has been produced in the context of the EU H2020 Project “GLASS - SinGLe Sign-on eGovernAnce paradigm based on a distributed file exchange network for Security, transparency, cost effectiveness and truSt”, which is co-funded by the European Commission under the Grant agreement ID: 959879. This publication reflects only the authors’ views and the Community is not liable for any use that may be made of the information contained therein.

%
%
\bibliographystyle{splncs04}
\bibliography{egov}

\end{document}